\documentclass[aps,prl,,showpacs,twocolumn,superscriptaddress]{revtex4-1}

\usepackage[pdftex]{graphicx} 
\usepackage{natbib}
\usepackage{booktabs}
\usepackage{color}

\newcommand{\one}{Fig.~\ref{f1}}
\newcommand{\two}{Fig.~\ref{f2}}
\newcommand{\three}{Fig.~\ref{f3}}

\newcommand{\Ir}{Sr$_2$IrO$_4$} 
\newcommand{\LaIr}{(Sr$_{1-x}$La$_x$)$_2$IrO$_4$} 
\newcommand{\Jone}{$J_{\rm{eff}}=1/2$}
\newcommand{\Jthree}{$J_{\rm{eff}}=3/2$}

\begin{document}
\title{Collapse of the Mott gap and emergence of a nodal liquid in lightly doped Sr$_2$IrO$_4$}

\author{A. de la Torre}
\affiliation{Department of Quantum Matter Physics, 24 Quai Ernest-Ansermet, 1211 Geneva 4, Switzerland}
\author{S. McKeown Walker}
\affiliation{Department of Quantum Matter Physics, 24 Quai Ernest-Ansermet, 1211 Geneva 4, Switzerland}
\author{F. Y. Bruno}
\affiliation{Department of Quantum Matter Physics, 24 Quai Ernest-Ansermet, 1211 Geneva 4, Switzerland}
\author{S. Ricco}
\affiliation{Department of Quantum Matter Physics, 24 Quai Ernest-Ansermet, 1211 Geneva 4, Switzerland}
\author{Z. Wang}
\affiliation{Swiss Light Source, Paul Scherrer Institut, CH-5232 Villigen PSI, Switzerland}
\affiliation{Department of Quantum Matter Physics, 24 Quai Ernest-Ansermet, 1211 Geneva 4, Switzerland}
\author{I. Gutierrez Lezama}
\affiliation{Department of Quantum Matter Physics, 24 Quai Ernest-Ansermet, 1211 Geneva 4, Switzerland}
\author{G. Scheerer}
\affiliation{Department of Quantum Matter Physics, 24 Quai Ernest-Ansermet, 1211 Geneva 4, Switzerland}
\author{G. Giriat}
\affiliation{Department of Quantum Matter Physics, 24 Quai Ernest-Ansermet, 1211 Geneva 4, Switzerland}
\author{D. Jaccard}
\affiliation{Department of Quantum Matter Physics, 24 Quai Ernest-Ansermet, 1211 Geneva 4, Switzerland}
\author{C. Berthod}
\affiliation{Department of Quantum Matter Physics, 24 Quai Ernest-Ansermet, 1211 Geneva 4, Switzerland}
\author{T. K. Kim}
\affiliation{Diamond Light Source, Harwell Campus, Didcot, United Kingdom}
\author{M. Hoesch}
\affiliation{Diamond Light Source, Harwell Campus, Didcot, United Kingdom}
\author{E. C. Hunter}
\affiliation{School of Physics and Astronomy, The University of Edinburgh, James Clerk Maxwell Building, Mayfield Road, Edinburgh EH9 2TT, United Kingdom}
\author{ R. S. Perry}
\affiliation{London Centre for Nanotechnology and UCL Centre for Materials Discovery, University College London, London WC1E 6BT, United Kingdom}
\author{A. Tamai}
\affiliation{Department of Quantum Matter Physics, 24 Quai Ernest-Ansermet, 1211 Geneva 4, Switzerland}
\author{F. Baumberger}
\affiliation{Department of Quantum Matter Physics, 24 Quai Ernest-Ansermet, 1211 Geneva 4, Switzerland}
\affiliation{Swiss Light Source, Paul Scherrer Institut, CH-5232 Villigen PSI, Switzerland}
\affiliation{SUPA, School of Physics and Astronomy, University of St Andrews, St Andrews, Fife KY16 9SS, United Kingdom}


\maketitle


{\bf 
Superconductivity in underdoped cuprates emerges from an unusual electronic state characterised by nodal quasiparticles and an antinodal pseudogap~\cite{Norman1998,Damascelli2003,Shen2005}. The relation between this state and superconductivity is intensely studied but remains controversial~\cite{Hanaguri2004,Kondo2009,Ghiringhelli2012,Comin2014,Kondo2011,Norman2005}.
The discrimination between competing theoretical models is hindered by a lack of electronic structure data from related doped Mott insulators. Here we report the doping evolution of the Heisenberg antiferromagnet {\Ir}, a close analogue to underdoped cuprates~\cite{Wang2011,Kim2012c,Watanabe2013,Kim2014}. We demonstrate that metallicity emerges from a rapid collapse of the Mott gap with doping, resulting in lens-like Fermi contours rather than disconnected Fermi arcs as observed in cuprates~\cite{Norman1998,Damascelli2003,Kondo2009,Shen2005,Comin2014}. Intriguingly though, the emerging electron liquid shows nodal quasiparticles with an antinodal pseudogap and thus bares strong similarities with underdoped cuprates. We conclude that anisotropic pseudogaps are a generic property of two-dimensional doped Mott insulators rather than a unique hallmark of cuprate high-temperature superconductivity.
}

The parent compounds of the cuprate high-temperature superconductors are characterised by a correlation induced excitation gap in a single half filled band and strong Heisenberg antiferromagnetic coupling of the spin moments.
Intriguingly, these characteristics are also found in {\Ir}, a layered 5$d$ transition metal oxide. {\Ir} is isostructural to La$_2$CuO$_4$ with planar IrO$_2$ layers forming a square lattice of Ir$^{4+}$ ions with a nominal $5d^5$ configuration. Strong spin-orbit interaction removes the orbital degeneracy of the $t_{2g}$ shell resulting in a single, half-filled band of spin-orbital entangled pseudospin {\Jone} states. Electron correlations cause a Mott-like insulating ground state~\cite{Kim2008,Martins2011} with $(\pi,\pi)$ antiferromagnetic ordering and ungapped spin-excitations with energies comparable to cuprates~\cite{Jackeli2009,Kim2012c}.

Besides these striking analogies, there are also notable differences from cuprates. 
The Coulomb repulsion in the Ir 5$d$ shell is weaker leading to comparable energy scales for the charge gap and spin excitation bandwidth~\cite{Fujiyama2012b}. 
Further, the non-interacting Fermi surface of {\Ir} is electron like and centered at $(0,0)$~\cite{Kim2008,Wang2011}, rather than hole-like as in cuprates~\cite{Damascelli2003}. Within a Hubbard model, this suggests a particle-hole conjugate doping phase diagram with a stronger tendency towards superconductivity in electron doped {\Ir}, as pointed out in Ref.~\cite{Wang2011}. A recent numerical study indeed found $d$-wave superconductivity in electron doped but not in hole doped {\Ir}~\cite{Watanabe2013}. The evolution of the microscopic electronic structure of {\Ir} with electron doping is therefore of considerable interest. 


%
%
\begin{figure*}[!ht]
\includegraphics[width=17.5cm]{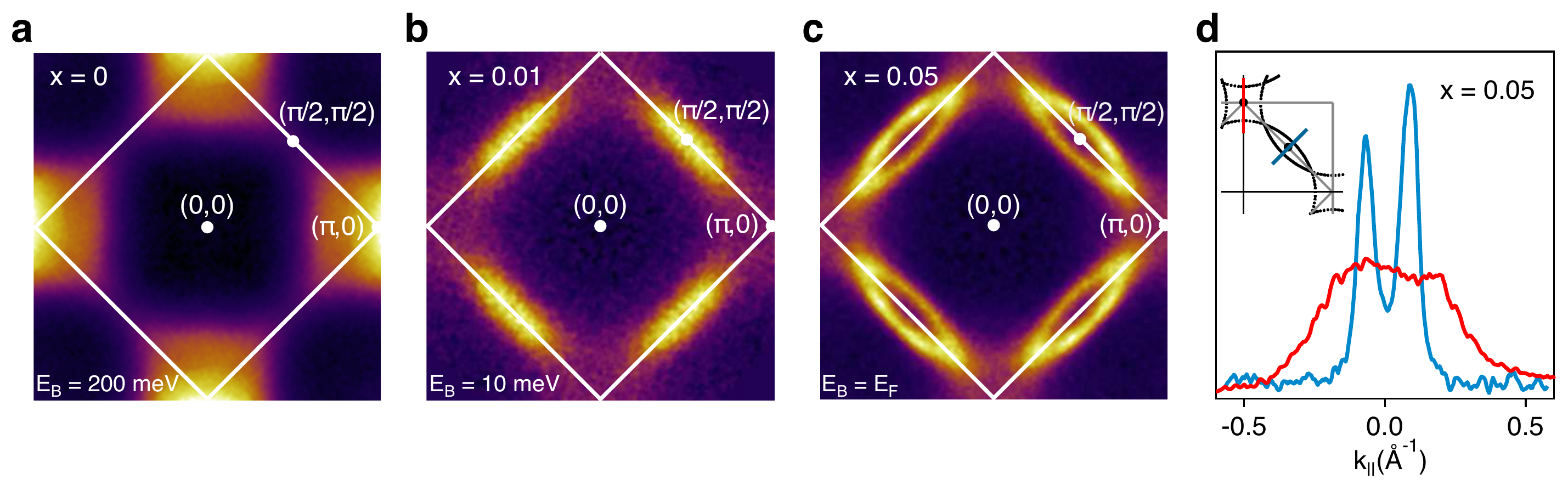} %
\caption{{\bf Nodal quasiparticles in lightly doped {\Ir}.} {\bf a}, Constant energy contour of {\Ir} at $-200$~meV showing the loci of the lowest lying charge excitations of the parent insulator in the large Brillouin zone corresponding to the Ir - Ir nearest neighbour square lattice. The white square illustrates the actual structural Brillouin zone that coincides with the magnetic zone and contains two iridium sites per plane.
{\bf b}, For $x=0.01$, the lowest lying spectral weight shifts from $(\pi,0)$ to $(\pi/2,\pi/2)$.
{\bf c}, At the higher doping of $x \approx 0.05$, we observe the emergence of quasiparticle like excitations along the nodal direction with clear spectral weight on the backside of the Fermi arcs. {\bf d}, Momentum Distribution Curves (MDCs) along the nodal (blue) and antinodal (red) direction illustrating the dichotomic behaviour of the single particle excitations. 
The data in {\bf a-c} has been four-fold rotationally averaged.}
\label{f1}
\end{figure*}

With this in mind, we investigated single crystals of {\LaIr} with $x=0$, 0.01 and 0.05 by angle resolved photoemission (ARPES). La$^{3+}$ substitutes for Sr$^{2+}$~\cite{Ge2011} and proved suitable to dope electrons in layered perovskites with minimal disorder induced scattering of in-plane carriers~\cite{Shen2007}. Crucially, La doping also preserves the strong spin-orbit interaction of Ir. A detailed characterization of the thermodynamic and transport properties of our samples is provided in Supplementary Information. For the highest doping of $x=0.05$, {\LaIr} shows a metallic resistivity down to $\sim 50$~K followed by an upturn at lower temperature, comparable to underdoped cuprates. We find no signs of superconductivity down to 100~mK. The magnetic ordering persists at $x=0.01$ with slightly reduced N\'eel temperature while samples with $x=0.05$ are paramagnetic. Note that because of the stoichiometry of {\Ir}, the nominal electron doping $x'$ on the Ir site is $2x$.

{\one} shows the evolution of the spectral weight near the Fermi level from the parent insulator $(x=0)$ to La concentrations of $x=0.01$ and 0.05. Consistent with an earlier report on undoped {\Ir}~\cite{Kim2008} we find the top of the lower Hubbard band (LHB) at the $(\pi,0)$ point. Yet, already for $x=0.01$ the low-energy spectral weight shifts to $(\pi/2,\pi/2)$. Increasing the doping to $x=0.05$, coherent quasiparticle-like excitations emerge along arcs stretching out from the nodal direction, while the spectral weight along the Ir-Ir nearest neighbour direction remains weak and is devoid of sharp features, reminiscent of the nodal-antinodal dichotomy in underdoped cuprates~\cite{Zhou2004,Shen2005}. However, in striking contrast to cuprates and to an earlier study on surface doped {\Ir}~\cite{Kim2014} we find strong spectral weight on the back side of the Fermi arcs. In order to understand this behaviour, it is important to recall the significant rotation of the octahedra in {\Ir}. This causes a purely structural $\sqrt{2} \times \sqrt{2}$ reconstruction of the IrO$_2$ plane and thus back folding into a small Brillouin zone that coincides with the magnetic zone~\cite{Kim2012}. Considering that equally strong back folding was observed previously in isostructural and non-magnetic Sr$_2$RhO$_4$~\cite{Baumberger2006} it is compelling to attribute the observation of lens-like low-energy contours, rather than isolated arcs in {\LaIr} to a purely structural effect. We thus turn our attention to the more fundamental issue of small versus large Fermi surface. 
A small Fermi surface comprising the doped carriers only was recently reported for the La doped bilayer iridate Sr$_3$Ir$_2$O$_7$, which shows characteristics of a correlated doped semiconductor and exhibits highly coherent quasiparticle states along the entire Fermi surface~\cite{DelaTorre2014}.
On the other hand, a large Fermi surface of volume $1+x'$ is observed in cuprates and is expected if all electrons in a doped single-band Mott insulator contribute to the Luttinger volume.
This question cannot be decided reliably based on the enclosed Fermi surface volume alone. 
For $x=0.05$, the difference between a large circular Fermi surface centered at $(0,0)$ and four lens-like small Fermi pockets at $(\pi/2,\pi/2)$ is minute and within the range of deviations from Luttinger's theorem observed in cuprates. It is thus essential to follow the high-energy evolution of the LHB with doping, which was not possible on surface doped {\Ir}~\cite{Kim2014}.
To this end, we compare in {\two} the band dispersion of {\LaIr} for $x=0$ and $x=0.05$.

%
%
\begin{figure*}[!ht]
\includegraphics[width=17cm]{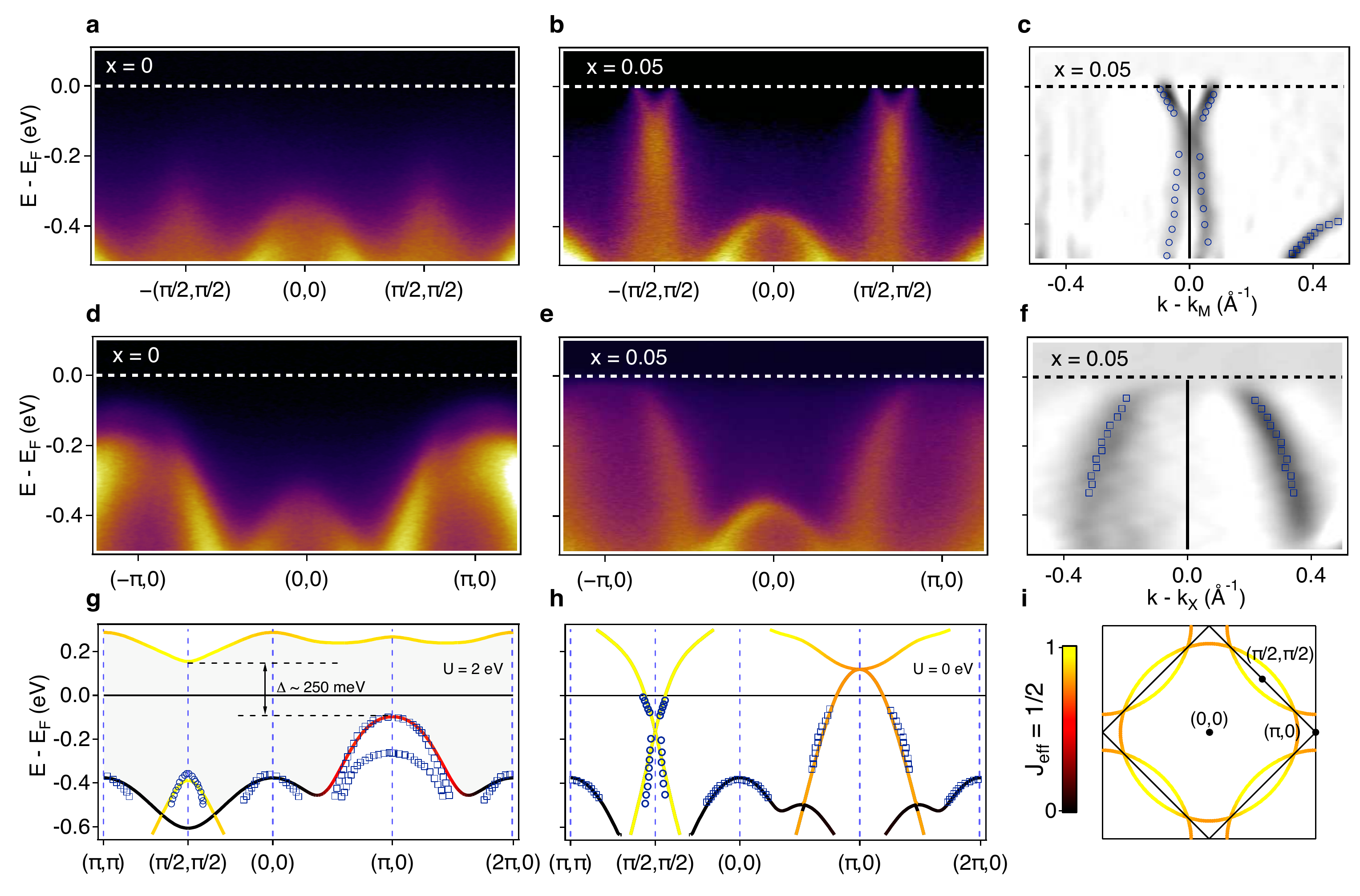} %
\caption{{\bf Collapse of the Mott gap}. {\bf a,d} Photoemission intensity in the fully gapped parent insulator along the nodal and antinodal direction, respectively. {\bf b,e} For $x=0.05$, electron like metallic states appear at $(\pi/2,\pi/2)$ while the apex of the hole-like band at $(\pi,0)$ moves above the chemical potential resulting in a collapse of the charge gap. {\bf c,f} Curvature plots of the raw data used to extract band positions.
{\bf g} Band dispersion for $x=0$ compared to a TB +SO + U calculation projected onto pseudospin {\Jone} states. {\bf h} Band dispersion for $x=0.05$ compared to a calculation with $U=0$. {\bf i} Large electron like Fermi surface corresponding to the TB calculation with $U=0$ and comprising $1+x'$ carriers. The color scale in {\bf g-i} encodes pseudospin {\Jone} character.}
\label{f2}
\end{figure*}

In undoped {\Ir} the electronic structure is dominated by gapped hole-like bands with maxima at $(0,0)$, $(\pi,0)$ and $(\pi/2,\pi/2)$ as shown by blue markers in {\two}~{\bf g}. For clarity, we describe the dispersion of these bands by a tight binding calculation that includes spin-orbit interaction and Coulomb repulsion (TB+SO+U)~\cite{Jin2009}. The latter is treated self-consistently in a mean field expansion of the density operator $n$ and thus cannot account for the different ways in which bands of different pseudospin characters are affected by electronic correlations. Despite the limitations of the model in handling correlations, we observe a good qualitative agreement between the calculated and the experimental dispersions for the parent compound using realistic parameters for the spin-orbit interaction ($\lambda=0.57$~eV) and Coulomb repulsion ($U=2$~eV). Full details of this model are given in the Supplementary Information. 
Projecting the wave functions onto a pseudospin basis we find {\Jone} states at $(\pi,0)$ and $(\pi/2,\pi/2)$ while the band with similar energy at $(0,0)$ is of {\Jthree} character.

Upon La doping, marked changes in the {\Jone} dispersion occur. Most notably, around $(\pi/2,\pi/2)$ bands with nearly linear dispersion appear at the Fermi level. 
These bands extrapolate to a Dirac point around $-0.1$~eV and continue to disperse quasi-linearly at higher energy, although we cannot presently exclude a small gap along the high-symmetry line. 
Along the antinodal direction, the hole-like {\Jone} band at $(\pi,0)$ shifts towards the chemical potential and clearly extrapolates to a band apex above the Fermi level ({\two}~{\bf e,f}). 
 The {\Jthree} bands, on the other hand, are not affected strongly by doping.
This behaviour is summarized in {\two}~{\bf h} where we show band positions for $x=0.05$ extracted from curvature plots of the raw data (see {\two}~{\bf c,f} and Supplementary Information).
Clearly, this electronic structure cannot be described by rigid shift of the tight binding band structure that describes the parent compound. Instead, we find that the band dispersion for $x=0.05$ is well approximated by a calculation with $U=0$
describing a weakly interacting metallic state. The low-energy excitations in doped iridates thus share a key-property of the first doped holes in cuprates, which also track the non-interacting band structure~\cite{Shen2004}. Even their nodal Fermi velocity of $\sim 10^5$~m/s is similar to lightly doped cuprates~\cite{Ino2002,Shen2005}.
However, the collapse of the Mott state is far more pronounced and rapid in iridates where already at $x=0.05$ no trace of the lower Hubbard band remains.
This is distinct from cuprates where the evolution from insulator to strange metal proceeds more gradually via a progressive transfer of spectral weight from the Hubbard band to the coherent quasiparticle band~\cite{Shen2004}.
While a profound understanding of this difference will require further theoretical work, we speculate that it reflects a reduced Mottness of {\Ir} arising from the weaker Coulomb repulsion in the 5$d$ shell.

%
%
\begin{figure*}[!ht]
\includegraphics[width=17cm]{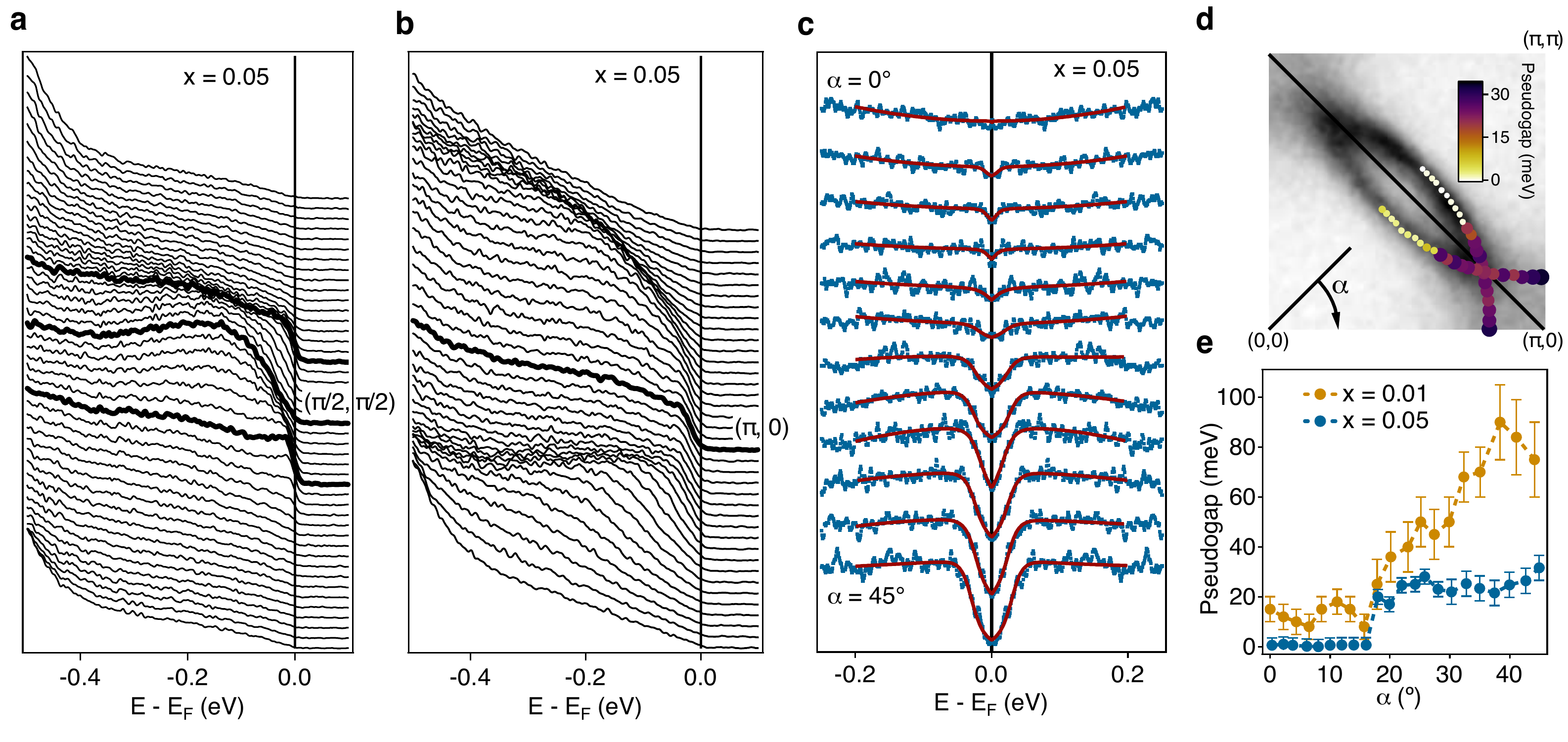} %
\caption{{\bf Anisotropic pseudogap for $x=0.05$.} {\bf a,b} Energy distribution curves along $(0,0)$ - $(\pi,\pi)$ and $(0,0)$ - $(\pi,0)$, respectively, showing a suppression of the low-energy spectral weight in the antinodal region (panel {\bf b}).
{\bf c} symmetrised EDCs (blue dots) and fits to the Dynes formula (red lines) for different angles along the Fermi surface. {\bf d} Fermi surface with the color-coded magnitude of the pseudogap overlaid. Note that the onset of the pseudogap is observed slightly before the apex of the lens-like contour.
{\bf e} Pseudogap as a function of Fermi surface angle for $x=0.01$ and $x=0.05$.}
\label{f3}
\end{figure*}

The gapless band structure found in doped {\Ir} corresponds to a large Fermi surface comprising $1+x'$ carriers, as shown in {\two} {\bf i}. This Fermi surface arises from {\Jone} states and is backfolded in the small Brillouin zone containing two Ir sites per plane giving the impression of lens-like Fermi pockets around $(\pi/2,\pi/2)$. However, these pockets are not separated by a significant gap from the square contours at $(\pi,0)$ that complete the large circular Fermi surface.
Hence, the suppression of the spectral weight at the antinode described in {\one} cannot be explained by a band gap opening between a lens-like small Fermi surface and and a fully occupied band at the $(\pi,0)$ point, as it is observed in electron doped Sr$_3$Ir$_2$O$_7$~\cite{DelaTorre2014}. Instead, it is indicative of a momentum dependent pseudogap. 

Direct evidence for a pseudogap is summarized in {\three}. Energy distribution curves along the node ({\three}~{\bf a}) show a sharp cutoff at the Fermi level, while the low-energy spectral weight is clearly suppressed along the antinode ({\three}~{\bf b}). In order to quantify the anisotropy of the pseudogap we fit particle-hole symmetrised EDCs along the large Fermi surface with the Dynes function~\cite{Dynes1978} often used to quantify superconducting gaps. While this procedure is purely phenomenological and cannot give absolute gap values, it is still suitable to monitor the evolution of the pseudogap with momentum. For $x=0.05$ where the spectral weight has a well defined cutoff along the entire Fermi surface we find no significant pseudogap near the node within the accuracy of the experiment of approximately 3~meV. Moving out along the Fermi surface towards the antinode, the pseudogap sets in suddenly at a Fermi surface angle of $\sim 17^{\circ}$. As shown in {\three}~{\bf d}, this angle is slightly smaller than the crossing with the Brillouin zone boundary suggesting that already the lens-like part of the Fermi surface is broken into two disconnected arcs and a small gapped region near the apex. 
Elucidating the precise doping range over which this behaviour exists will require further detailed measurements. For $x=0.01$, the suppression of spectral weight has a less clearly defined onset in energy. However, consistent with the insulating nature of this sample, a pseudogap is clearly present along the entire Fermi surface and reaches values up to $\sim 80$~meV near the antinode ({\three}~{\bf e} and Supplementary Figure 3).

The origin of the pseudogap in {\Ir} cannot be determined unambiguously from our present data.
In cuprates, a pseudogap with strikingly similar phenomenology is often associated with preformed non-phase coherent pairs~\cite{Norman1998,Kondo2011} or competing ordered states~\cite{Comin2014}. Yet, the pseudogap in {\Ir} persist above 100~K, while our samples show no superconductivity down to 100~mK at a doping level of $x'=0.1$ where hole doped cuprates are superconducting. 
Moreover, neither our ARPES data nor diffraction experiments~\cite{Kim2012,Fujiyama2012b} give evidence of competing ordered states as they are found by different techniques in the pseudogap phase of cuprates~\cite{Ghiringhelli2012,Hanaguri2004,Comin2014}.
Taken together with the similarities of the magnetic excitations in the parent compounds, our results thus suggest that an anisotropic pseudogap is an intrinsic property of lightly doped low-dimensional Mott insulators with Heisenberg spin-dynamics. This possibility is consistent with dynamical mean field theory studies of underdoped cuprates~\cite{Huscroft2001,Parcollet2004,Ferrero2009}.

{\bf Methods}
Crystals of {\LaIr} were flux grown by heating a mixture of off-stoichiometric quantities of IrO$_2$, La$_2$O$_3$ and SrCO$_3$ in an anhydrous SrCl$_2$ flux. The resulting crystals, ranging in size from 200~$\mu$m to 600~$\mu$m, were mechanically separated from the flux by washing with water. The La concentration $x$ was determined by energy dispersive x-ray spectroscopy in each sample measured by ARPES. The crystals were further characterised by resistivity, magnetic susceptibility, heat capacity and both powder and single crystal x-ray diffraction measurements. Details of the sample growth and characterization are provided in the Supplementary Information.

ARPES measurements were performed at the I05 beamline of the Diamond Light Source. The samples were cleaved at pressures $<10^{-10}$~mbar and temperatures $<50$~K. Measurements were made using photon energies between 30~eV and 120~eV. All presented data were acquired at 100~eV with an energy resolution of 15~meV. The sample temperature was 8~K and 50~K for conducting and insulating samples, respectively. 

\bibliography{Ir_bib_v4.bib}

\end{document}